\documentclass[conference]{IEEEtran}
\IEEEoverridecommandlockouts
\usepackage{cite}
\usepackage{multirow}
\usepackage{pifont}
 \usepackage{amsmath,amsfonts,amsthm,bm} 
\usepackage{amsmath,amssymb,amsfonts}
\usepackage{algorithmic}
\usepackage{graphicx}
\usepackage{float}
\usepackage{multirow}
\usepackage{multicol}
\usepackage{epstopdf}
\usepackage{textcomp}
\usepackage{xcolor}
\usepackage{stackengine}
\usepackage{algorithm}
\usepackage{bbding}

\begin{document}

\title{ \large Content Caching-Assisted Vehicular Edge Computing Using Multi-Agent Graph Attention Reinforcement Learning}

\author{Jinjin Shen, 
Yan Lin,~\IEEEmembership{Member,~IEEE}, 
Yijin Zhang,~\IEEEmembership{Senior Member,~IEEE},
Weibin Zhang,~\IEEEmembership{Member,~IEEE}, ~\\
Feng Shu,~\IEEEmembership{Member,~IEEE} and 
Jun Li,~\IEEEmembership{Senior Member,~IEEE}

\thanks{Copyright (c) 20xx IEEE. Personal use of this material is permitted. However, permission to use this material for any other purposes must be obtained from the IEEE by sending a request to pubs-permissions@ieee.org. This work was supported in part by the National Natural Science Foundation of China under Grants 62001225, 62071236, 62471204, and U22A2002, in part by Hainan Province Science and Technology Special Fund under Grant ZDYF2024GXJS292, in part by the Scientific Research Fund Project of Hainan University under Grant KYQD(ZR)-21008, in part by the Collaborative Innovation Center of Information Technology, Hainan University, under Grant XTCX2022XXC, in part by the Key
Technologies R\&D Program of Jiangsu (Prospective and Key Technologies for Industry) under Grants BE2023022 and BE2023022-2, in part by Natural Science Foundation of Jiangsu Province under Grant BK2021022532, and in part by Major Natural Science Foundation of the Higher Education Institutions of Jiangsu Province under Grant 24KJA510003.
 (\textit{Corresponding author: Yan Lin)}

J. Shen, Y. Lin, W. Zhang and Y. Zhang are with the School of Electronic and Optical Engineering, Nanjing University of Science and Technology, Nanjing 210094, China (e-mail:\{jinjin.shen, yanlin, weibin.zhang\}@njust.edu.cn;
yijin.zhang@gmail.com). F. Shu is with the School of Information and Communication Engineering, Hainan University, Haikou 570228, China (email: shufeng0101@163.com). J. Li is with the School of Information Science and Engineering, Southeast University, Nanjing 210096, China (email: jleesr80@gmail.com).}}

\maketitle
\begin{abstract}
In order to avoid repeated task offloading and realize the reuse of popular task computing results, we construct a novel content caching-assisted vehicular edge computing (VEC) framework. In the face of irregular network topology and unknown environmental dynamics, we further propose a multi-agent graph attention reinforcement learning (MGARL) based edge caching scheme, which utilizes the graph attention convolution kernel to integrate the neighboring nodes' features of each agent and further enhance the cooperation among agents. Our simulation results show that our proposed scheme is capable of improving the utilization of caching resources while reducing the long-term task computing latency compared to the baselines.
\end{abstract}

%\vspace{-0.2cm}
\section{Introduction}
\vspace{-0.05cm}
Recently, mobile edge computing (MEC) has evolved as a promising technology for ultra-reliable and ultra-low latency (URLLC) applications, such as virtual reality and autonomous driving~\cite{autonomousdriving}. Instead of sending tasks to the far-end cloud, MEC sinks computing and caching resources to the edge nodes close to the users~\cite{closetotheusers}. In vehicular networks, vehicles have limited computing capability but can offload requested tasks to edge servers (ESs)
%e.g. the surrounding Roadside Units (RSUs) or Base Stations (BSs), 
for execution~\cite{forexecution1,forexecution2,forexecution3}, thus relieving the computing pressure of vehicles~\cite{relievecomputingpressure}.

Nevertheless, in practical scenarios where vehicular users (VUs) are driving during rush hour traffic or in congested urban areas, the data they offload exhibits high spatial, temporal, and social correlation, which results in different VUs requesting the same services simultaneously, such as navigation assistance, real-time traffic updates, or recommendations for nearby amenities~\cite{nearbyamenities}. As such, multiple VUs may have similar computing tasks associated with the same computing results to access and utilize the shared services. In this case, instead of duplicate task re-uploading and re-computing, the previous task computing results can be cached by ESs, e.g. the surrounding Roadside Units (RSUs) or VUs, and be further utilized by other VUs to reduce the latency of subsequent tasks~\cite{tasksarewasted}. Therefore, designing an optimal content caching policy for edge computing is of great significance for the vehicular edge computing (VEC) networks.

Considering the unknown dynamics of time-varying network topology and
channel conditions, the edge computing content caching problem is essentially a model-free sequential decision-making problem~\cite{decisionmakingproblem}. 
By combining the advantages of deep learning in identifying data features and reinforcement learning in dynamic programming, deep reinforcement learning (DRL) has been widely employed in solving such problems~\cite{Geng}. 
For example, H. Tian \textit{et al.} of~\cite{copace} used deep deterministic policy gradient (DDPG) approach to orchestrate the joint offloading, caching, and resource allocation decision-making for VEC, taking into account the time-varying content popularity. However, it relies on the centralized learning framework without cooperative learning.  Additionally, X. Ye \textit{et al. }of ~\cite{yexinyu} designed a blockchain-based collaborative computing
and caching framework in VEC using the multi-agent asynchronous advantage actor-critic (A3C) approach, which has verified the efficiency of cooperative learning compared to the non-cooperative learning.

%and the content reuse is not utilized. 
 %However, the centralized learning framework was employed and the irregular network topology characteristic was neglected. 

 %As a further step, the authors of [11] investigated the joint caching, communication, and computing resource allocation problem in MEC by taking into account the content popularity and graphical information of ESs using multi-agent deep graph reinforcement learning approach for cooperative learning. However, the proposed scheme is only suitable for static MEC networks.

\begin{table}[t]
\centering
\renewcommand{\arraystretch}{1.2}
\caption{{Related contributions}}
\begin{tabular}{|l|ll|l|l|l|}
\hline
\multirow{2}{*}{\textbf{}} & \multicolumn{2}{c|}{Network Topology}                 & \multicolumn{1}{c|}{\multirow{2}{*}{\begin{tabular}[c]{@{}c@{}}Content \\ Popularity\end{tabular}}} & \multicolumn{1}{c|}{\multirow{2}{*}{\begin{tabular}[c]{@{}c@{}}Graphical \\Information  \end{tabular}}} & \multicolumn{1}{c|}{\multirow{2}{*}{\begin{tabular}[c]{@{}c@{}}Cooperative\\  Learning\end{tabular}}} \\ \cline{2-3}
 & \multicolumn{1}{l|}{Dynamic} & Irregular              &                                     &                                &                                         \\ \hline
\multicolumn{1}{|c|}{{[}10{]}}   & \multicolumn{1}{c|}{\checkmark}       & \multicolumn{1}{c|}{\ding{55}} & \multicolumn{1}{c|}{\ding{55}}   &   \multicolumn{1}{c|}{\ding{55}} &  \multicolumn{1}{c|}{\checkmark}  \\ \hline

\multicolumn{1}{|c|}{{[}11{]}}  & \multicolumn{1}{c|}{\checkmark} &  \multicolumn{1}{c|}{\ding{55}}  &  \multicolumn{1}{c|}{\checkmark} &  \multicolumn{1}{c|}{\ding{55}} &  \multicolumn{1}{c|}{\ding{55}}\\ \hline
\multicolumn{1}{|c|}{{[}12{]}}                   & \multicolumn{1}{c|}{\checkmark}        & \multicolumn{1}{c|}{\ding{55}}  & \multicolumn{1}{c|}{\checkmark}  & \multicolumn{1}{c|}{\ding{55}}  & \multicolumn{1}{c|}{\checkmark} \\ \hline
\multicolumn{1}{|c|}{{[}16{]}}                   & \multicolumn{1}{c|}{\ding{55}}        & \multicolumn{1}{c|}{\ding{55}}  &  \multicolumn{1}{c|}{\checkmark}  &   \multicolumn{1}{c|}{\checkmark}   &  \multicolumn{1}{c|}{\checkmark}  \\ \hline
{}\textbf{Ours}{}                   & \multicolumn{1}{c|}{\checkmark}        & \multicolumn{1}{c|}{\checkmark}  &  \multicolumn{1}{c|}{\checkmark}  &   \multicolumn{1}{c|}{\checkmark}   &  \multicolumn{1}{c|}{\checkmark}  \\ \hline

\end{tabular}
\label{compare}
\vspace{-0.4cm}
\end{table}

However, the randomness of VU mobility leads to the dynamics and temporal-spatial irregularities of the network
topology, which results in the environmental uncertainty and makes environmental knowledge more difficult to learn.
But the aforementioned commonly used DRL solutions relying on convolutional neural networks exhibit weakness in extracting the
temporal and spatial relationships between nodes in the irregular topology. To solve this issue, graph convolutional neural networks have been employed in DRL, which has the benefits of acting directly on graphs and fully utilizing the structural information such as the relationships among nodes~\cite{survey}. Consequently, it has been used for solving optimization problems by jointly considering the information of agents close to an agent~\cite{DGN}, and further enhances the cooperation of agents~\cite{cooperationofagents}. Although D. Wang \textit{et al. }of~\cite{suchproblems} have explored the deep graph reinforcement learning approach for solving the joint caching, communication, and resource allocation problem, they did not carefully consider the dynamic graphical topology characteristics resulted from high vehicular mobility. In this context, this paper aims to investigate the content caching-assisted VEC problem, where the irregular and dynamic network topology and the cooperative learning among multiple vehicles are considered. \textit{To the best of our knowledge, at the time of writing, this is the first attempt in the related literature to study the content caching problem in VEC networks relying on multi-agent graph attention reinforcement learning (MGARL) framework.} Compared to the existing literature, MGARL method has the promising potential to capture the dynamics and irregularities of time-varying vehicular networks, as well as make better use of the neighboring nodes' information to facilitate the cooperative content caching decision-making.

Our main contributions are boldly and explicitly contrasted to the literature in Table~\ref{compare} and are detailed as follows:
%\vspace{-0.05cm}
\begin{itemize}
\item A content caching-assisted VEC framework is established where each VU needs to decide whether to cache the task computing result, with the aim of reducing computing latency and reusing the popular content. 
\item The edge caching decision-making problem is constructed as a decentralized partially observable Markov Decision Process (DEC-POMDP), where each VU agent observes its local information and takes action individually based on its learned policy. 
\item An MGARL-based edge caching scheme is proposed, where graph attention convolution kernels are utilized to capture relationships among agents by taking into account the neighboring agents' features when making their own edge caching decisions.
\item Simulation results show that the proposed scheme outperforms other baselines in terms of both improving caching resource utilization and reducing long-term task computing latency.
%\vspace{-0.15cm}
\end{itemize}

\section{System Model and Problem Formulation}
\vspace{-0.05cm}
\subsection{Network Model}

\begin{figure}
\vspace{-0.5cm}
\centering 
\includegraphics[width=0.8\linewidth]{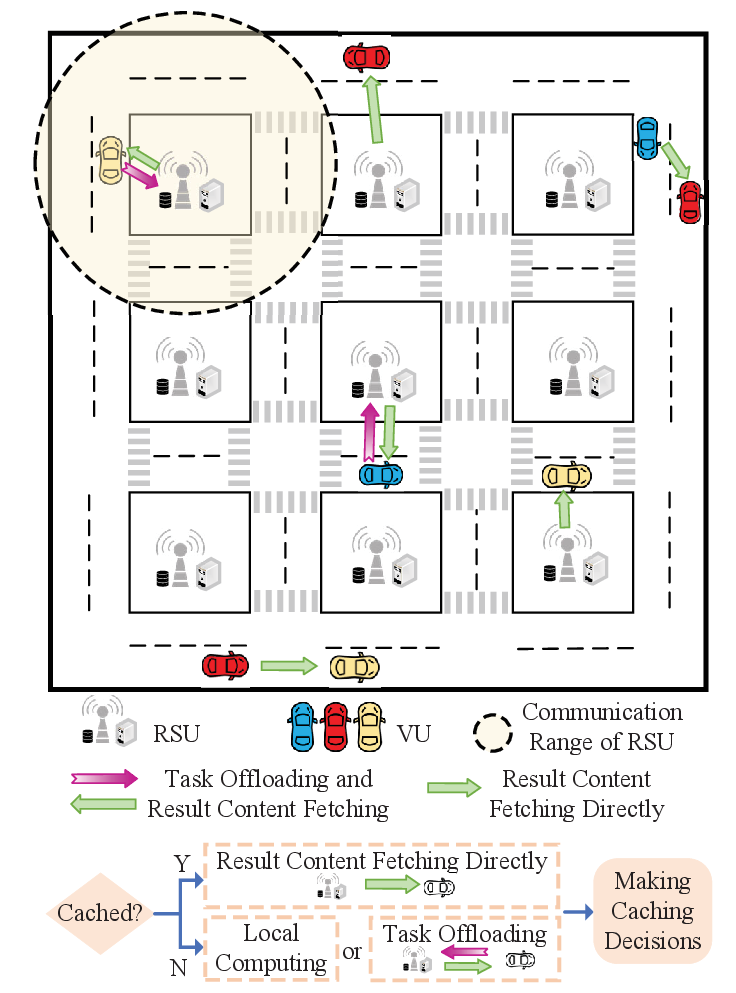}
\vspace{-0.3cm}
    \caption{Illustration of the content caching-assisted VEC system.}
    \label{model}
\vspace{-0.3cm}
\end{figure}

\vspace{-0.05cm}
As shown in Fig.~\ref{model}, we consider a Manhattan vehicular network model which consists of multiple horizontal and vertical two-lane two-way roads. $L$ RSUs are deployed uniformly and $M$ VUs drive along the road with their velocity obeying the Markov-Gaussian stochastic process independently and may change their direction at intersections with a given probability $\eta$. Let $ \mathcal{L}=\{1,2,\cdots,L\}$  and $\mathcal{M}=\{1,2,\cdots,M\}$ denote the index sets of RSUs and VUs, respectively. Both RSUs and VUs have the capability to compute tasks and cache contents, which are referred to as service nodes (SNs). Let $ \mathcal{N}=\mathcal{L} \cup \mathcal{M}=\{1,2,\cdots,N\}$ represent the index set of $N$ SNs.

Without loss of generality, we assume that the system operates in a time-slot (TS) mode and environmental parameters (e.g. transmit power and channel gain) remain unchanged during each TS. Additionally, new tasks with different popularity characteristics arrive every certain TSs, and each task can be completed successfully during each TS. Let $\mathcal{K}=\{1,2,\cdots,K\}$ represent the index set of tasks, where task $k$ is represented by a 3-tuple $ \left\langle d_k,s_k,b_k \right\rangle$ with the task size $d_k$, the number of CPU cycles required to process the task $s_k$ and the size of the processed task content $b_k$. Note that, if the content of the task result has been cached on SNs within the communication range of the VU, the content can be fetched directly from SNs to VUs.
Otherwise, the VU has to compute the task itself or offload it for execution. In this case, after completing the task, each VU needs to decide whether to cache the content and to which SN. 
\vspace{-0.1cm}
\subsection{Communication Model}
\vspace{-0.05cm}
Let $P_m$ and $g_{m,n}^{\text{up}}(t)$ denote the transmit power of VU $m$ and the channel gain spanning from VU $m$ and SN $n$ in TS~$t$, respectively. In this framework, we consider the small-scale fading and the path loss of vehicle-to-vehicle (V2V) communication and vehicle-to-infrastructure (V2I) communication, neglecting the effects of shadow fading~\cite{forexecution2}. We assume that the inter-user interference has been eliminated by allocating orthogonal resource blocks. Thus, the uplink transmission rate from VU $m$ to SN $n$ in TS $t$ is given by 
%\vspace{-0.2cm}
\begin{equation}
\zeta_{m,n}^{\text{up}}(t)=B\log_2{(1+{P_mg_{m,n}^{\text{up}}(t)}/{\sigma^2})}
\label{up},
%\vspace{-0.2cm}
\end{equation}
where $B$ is the bandwidth of each channel, and $\sigma^2$ is the power of the additive Gaussian white noise, both of which are assumed to be the same for each TS. Similarly, let $P_n$ and $g_{n,m}^{\text{down}}(t)$ denote the transmit power of SN $n$ and the channel gain spanning from SN $n$ and VU $m$ in TS~$t$, respectively. Thus the downlink transmission rate from SN $n$ to VU $m$ in TS $t$ is given by
%\vspace{-0.2cm}
\begin{equation}
\zeta_{n,m}^{\text{down}}(t)=B\log_2{(1+{P_ng_{n,m}^{\text{down}}(t)}/{\sigma^2})}
\label{down}.
\end{equation}

\vspace{-0.3cm}
\subsection{Computing Model}
\vspace{-0.05cm}
Let $k_m(t)$ denote the task requested by VU $m$ in TS $t$ and $w_m(t)\in \{0,1\}$ 
 denote whether VU $m$  has to offload $k_m(t)$.
 %or compute $k_m(t)$ itself in TS $t$. 
 If $w_m(t)=0 $, VU $m$ computes $k_m(t)$ itself, otherwise VU $m$ offloads $k_m(t)$ to the nearest SN $n_{m}(t)$. 
 
\begin{itemize}
\item\textbf{Local Computing:}  Let $f_m^{\text{l}}$ be the computation capability on VU $m$. Therefore, the latency of computing task $k_m(t)$ generated by user VU $m$ is given by $T_m^{\text{l}}(t)=(1-w_m(t)){s_{k_m(t)}} / {f_m^{\text{l}}}$.

\item\textbf{Task Offloading:}
Let $f_{n_m(t)}^{\text{off}}$ be the computation capability on SN $n_m(t)$. Then, the latency in completing task $k_m(t)$ is expressed as $T_{m}^{\text{off}}(t)=w_m(t)(T_{m}^{\text{up}}(t)+T_{m}^{\text{exe}}(t)+ T_{m}^{\text{down}}(t))$, where $T_{m}^{\text{up}}(t)={d_{k_m(t)}} / {\zeta_{m,n_m(t)}^{\text{up}}(t)}$, $T_{m}^{\text{exe}}(t)={s_{k_m(t)}} / {f_{n_m(t)}^{\text{off}}}$ and $T_{m}^{\text{down}}(t)={b_{k_m(t)}} / {\zeta_{n_m(t),m}^{\text{down}}(t)}$ denote the task offloading latency, task computing latency, and result content fetching latency, respectively.
%\vspace{-0.2cm}
\end{itemize}
\vspace{-0.05cm}
\subsection{Caching Model}
\vspace{-0.05cm}
Let $\textbf{H}_{\text{ca}}(t)=[h_{n,k}^{\text{ca}}(t)]_{\forall n,\forall k}$ represent whether computing result contents are cached or not by all SNs in TS~$t$. Specifically, $h_{n,k}^{\text{ca}}(t)=1$ if the content of task $k$ is cached on SN $n$ in TS $t$, otherwise $h_{n,k}^{\text{ca}}(t)=0$. Note that all SNs have limited caching capacity with $g_n(t) $ for SN $n$, thus it is impossible to cache all the task contents. Let us continue to denote $a_m^{\text{ca}}(t)\in\{0,1,\cdots,N\}$ as the caching decision variable for VU $m$ in TS $t$. To be specific, if $a_m^{\text{ca}}(t)=n~(n\neq0)$, VU $m$ caches the computing result content to SN $n$ and $h_{n,k}^{\text{ca}}(t)=1$, otherwise VU $m $ does not cache the content to any SN. Then, the caching decision of all VUs in TS $t$ can be denoted as $\textbf{a}_{\text{ca}}(t)=[a_1^{\text{ca}}(t),a_2^{\text{ca}}(t),\cdots,a_M^{\text{ca}}(t)]$.

When SNs cache the content of $Z_t$ task computing results in TS $t$, we have $\mathcal Z=\{1,2,\cdots,Z_t\}$. We assume that the task content popularity follows the Zipf distribution, thus the probability of task content $z\in \mathcal{Z}$ to be requested is expressed as $q_z(t)={I_z(t)^{-\delta}} / {\sum_{z=0}^ZI_z(t)^{-\delta}}$
, where $I_z(t)$ denotes the ranking of the popularity of task content $z$ in descending order in TS~$t$, and $\delta$ is the Zipf distribution parameter.
\vspace{-0.05cm}
\subsection{Problem Formulation}
\vspace{-0.05cm}
Our aim is to design an edge caching scheme for the VEC system to minimize the long-term task computing latency by utilizing limited edge caching resources. To formulate our problem, we introduce an auxiliary variable $e_m(t)$, where $e_m(t)=1$ when $m$ can fetch the required content directly from SNs within its communication range, otherwise $e_m(t)=0$. Thus, the total computation latency is the local computing latency or the task offloading latency when $e_m(t)=0$, and is the latency for fetching the cached task result when $e_m(t)=1$. Then, the task computing latency for VU $m$ in TS $t$ can be expressed by $T_m(t)=(1-e_m(t))(T_{m}^\text{l}(t)+T_{m}^{\text{off}}(t))+e_m(t)T_{m}^{\text{down}}(t)$. Mathematically, the optimization problem is formulated as
\begin{subequations}
\vspace{-0.15cm}
\begin{equation}\min_{\left \{\textbf{a}_{\text{ca}}(t) \right \}}{E[\sum_{t=1}^T{\sum_{m=1}^{M}}T_m(t) ]}\end{equation}
\vspace{-0.2cm}
\begin{equation}\text{s.t.}\:~a_m^{\text{ca}}(t)\in \left \{ 0,1,\cdots ,N\right \} ,\forall m,\forall t,\end{equation}
\vspace{-0.5cm}
\begin{equation}\sum_{k\in\mathcal{K}}h_{n,k}^{\text{ca}}(t)b_k\le g_n(t),\forall n,\label{c2}\end{equation}
\vspace{-0.3cm}
\end{subequations}

\noindent where constraint (\ref{c2}) indicates that the occupied caching capacity at SN $n$ can not exceed its maximum capacity $g_n(t)$.

\begin{figure*}[h]
\vspace{-0.4cm}
\centering
\includegraphics[width=0.79\textwidth,height=0.35\textwidth]{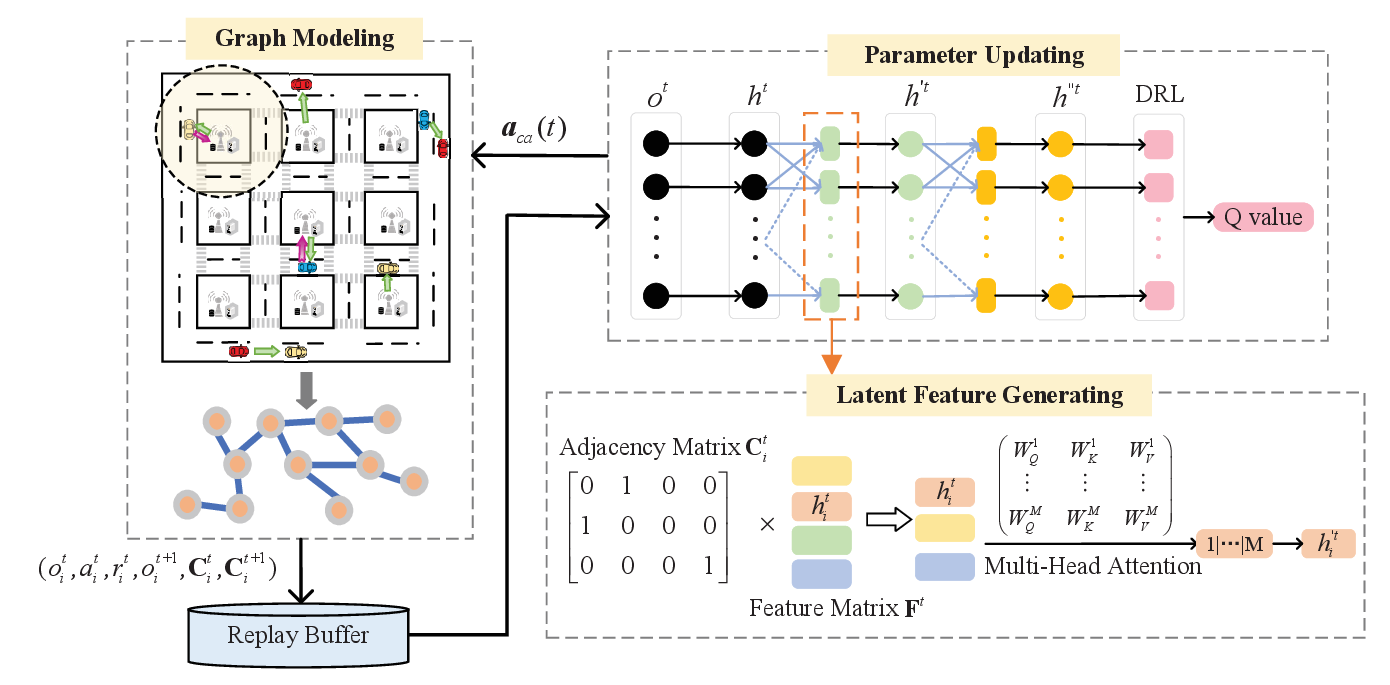}
\vspace{-0.4cm}
\caption{Illustration of the proposed MGARL-based content caching-assisted VEC scheme.}
\label{yuanli}
\vspace{-0.3cm}
\end{figure*} 

\section{The Design of DEC-POMDP}
\vspace{-0.05cm}
Considering the fact that the vehicular environment is dynamically changing and each VU is unable to obtain global environmental knowledge, we further formulate the above edge caching decision-making problem as a DEC-POMDP. Explicitly, each VU agent gets the local observation and takes action individually. By interacting with the environment iteratively, the agents can learn their strategies to maximize the system reward. Next, we will elaborate on the definitions of DEC-POMDP.
\vspace{-0.1cm}
\subsection{Observation}\label{AA}
\vspace{-0.1cm}
%Let $S$ denote the state space of the system, which consists of the horizontal and vertical coordinates of all SNs, the caching state of all SNs and the remaining storage state of all SNs. %Then the system state in time slot $t$ is denoted as
%\begin{equation}
%\begin{aligned}
%\pmb{s}_t=[x_1(t),x_2(t),\cdots,x_N(t),y_1(t),y_2(t),\cdots,y_N(t),\\h_{1,1}^{\text{ca}}(t),h_{1,2}^{\text{ca}}(t),\cdots,h_{N,K}^{\text{ca}}(t),g_1(t),g_2(t),\cdots,g_N(t)],
 %\label{eq13}
 %\end{aligned}
%\end{equation}

%\noindent where $x_n(t)$ and $y_n(t)$ are the current horizontal and vertical coordinates of SN $w_n$ respectively. 
Due to limited sensing and positioning technology, we assume that each VU agent can only observe its location, its caching state, and the current remaining caching capacity of all SNs. Accordingly, the observation of VU agent $m$ can be defined as 
\begin{equation}
\begin{aligned}
\pmb{o}_m(t)=[x_m(t),y_m(t),h_{m,1}^{\text{ca}}(t),h_{m,2}^{\text{ca}}(t),\cdots,\\h_{m,K}^{\text{ca}}(t),g_1(t),g_2(t),\cdots,g_N(t)],
 \label{eq14}
  \end{aligned}
\vspace{-0.2cm}
\end{equation}
\noindent where $x_m(t)$ and $y_m(t)$ are the current horizontal and vertical coordinates of VU agent $m$, respectively.
\vspace{-0.15cm}
\subsection{Action}
%\vspace{-0.15cm}
Based on the learned policy and its observation in TS $t$, VU agent $m$ selects an action $a_m^{ca}(t)$ to decide whether to cache the content and to which SN. 
%In TS $t$, the action selected by all VUs is represented as $\textbf{a}_{\text{ca}}(t)$.

\subsection{Reward Function}
%\vspace{-0.05cm}
In order to minimize the long-term task computing latency, the local reward of VU agent $m$ can be designed as
\begin{equation}
r_m(t)=e_m(t)re_m-T_m(t)+pl(t),
\label{reward}
\end{equation}
\noindent where $re_m$ is the reward for encouraging caching computing results. Moreover, $pl(t)$ is negative if the caching decision made by VU agent $m$ leads to the excess of caching capacity, otherwise $pl(t)=0$. Accordingly, the system reward, which is defined as the sum of all local rewards, is given by $r(t)=\sum_{m=1}^Mr_m(t)$.

\vspace{-0.1cm}
\section{MGARL-based Scheme}
\vspace{-0.1cm}
To adapt to dynamically changing irregular topology and capture relationships among agents, we resort to MGARL to further utilize the cooperation of multiple agents for solving the edge caching decision-making problem. As illustrated in Fig.\ref{yuanli}, MGARL consists of three modules: graph modeling, latent feature generating and parameter updating.

\subsection{Graph Modeling}

The vehicular network topology can be modeled as a graph $\mathcal{G}=\{\mathcal{V},\mathcal{E}\}$ where the node set $\mathcal{V}$ is the set of $M$ VUs and the edge set $\mathcal{E}$ is determined by the distance among vehicles. To be specific, there exists an edge between two nodes when their distance does not exceed the maximum communication range and such vehicles are neighbors of each other. Let $\mathcal{A}_i^t$ denote the set of neighbors of node $i$ in TS $t$. Each node in the graph has its node features, denoted by $h_i^t$ for node $i$ in TS $t$, which are yielded from the observation $o_i^t$ through a fully connected network.

\vspace{0.3em}
\begin{algorithm}
\vspace{0.3em}
\noindent \textbf{Algorithm 1} Training Process for MGARL-Based Content Caching-Assisted VEC Scheme
\vspace{0.3em}
\hrule
	\label{alg:1}
	\begin{algorithmic}[1]
 \vspace{0.1em}
		\FOR{each agent $i\in \mathcal{V}$}
		\STATE Initialize the $Q$ network parameter $\theta_i$ of agent $i$; 
		\ENDFOR 
		\FOR{each episode $q\in\{1,2,\cdots,Q\}$}
        \STATE Reset simulation parameters;
		\FOR{each TS $t\in\{1,2,\cdots,T\}$}
		\FOR{each agent $i\in \mathcal{V}$}
        \STATE
        VU $i$ completes one task;
        %successfully by fetching the result content from SNs, task offloading, or local computing;
        \STATE 
        Get local observation $o_i^t$ and construct the adjacency matrix $\mathbf{C}_i^t$;
		\STATE Select $a_i^t$ based on the learned policy;
            \STATE Cache the content and calculate  $r_i^t$;
            %by Eq.\ref{reward};
            \STATE Get $ o_i^{t+1} $ and $ \mathbf{C}_i^{t+1} $;
            \STATE Store the tuple $ \left\langle o_i^t,a_i^t,r_i^t,o_i^{t+1}, \mathbf{C}_i^t, \mathbf{C}_i^{t+1}\right\rangle $ in shared replay buffer;
            \STATE Encode $o_i^t$ to $h_i^t$ through a fully connected network; 
            \STATE Integrate the features of $\mathcal{A}_{+i}^t$  according to $\mathbf{C}_i^t$ to generate $h_i^{\prime t}$ by adopting multi-head attention as the convolution kernel;
            \STATE Sample a random minibatch of $S$ tuples from the shared replay buffer;
            \STATE Update $\theta_i$ to minimize the loss function;
            %by Eq.\ref{eq12};
		\ENDFOR
  \ENDFOR
  \ENDFOR
  \vspace{0.1em}
  \end{algorithmic}
\end{algorithm}

\vspace{-0.2cm}

\subsection{Latent Feature Generating}
The latent feature generating stage utilizes the convolutional layer to integrate the node features in node $i$'s local region, which includes $i$ and $\mathcal{A}_i^t$ to generate the latent feature $h_i^{\prime t}$ in TS~$t$. Firstly, in TS~$t$, all nodes’ feature vectors are merged into a feature matrix $\mathbf{F}^t$ with size $M \times D$ in the order of index where $D$ is the length of the feature vector. Then, let $\mathcal{A}_{+i}^t$ represent the set of node $i$ and $\mathcal{A}_i^t$ and we introduce an adjacency matrix $\mathbf{C}_i^t$ with size $\lvert \mathcal{A}_{+i}^t\rvert \times M$ to denote the one-hot representations of $\mathcal{A}_{+i}^t$. To be specific, the first row of $\mathbf{C}_i^t$ is the one-hot representation of node $i$, and the $j\in\{2,3,\cdots,\lvert \mathcal{A}_{+i}^t\rvert\}$th row is the representation of the $(j-1)$th neighbor of $i$. Then, the features in the local region of node $i$ are obtained by  $\mathbf{C}_i^t\times \mathbf{F}^t $.

To further capture the relationships among nodes, the multi-head attention is adopted as the convolution kernel to integrate the feature vectors in the local region of node $i$ and generate the latent feature vector $h_i^{\prime t}$. 
%The input feature of each node is projected to query, key and value representation. 
Let $W_Q^u$, $W_K^u$ and $W_V^u$ denote the parameter matrices corresponding to the query, key, and value representation in attention head $u$, respectively. For attention head $u$, the relationship between node $i$ and its neighbor $j\in \mathcal{A}_{+i}^t$ in TS $t$ can be formulated as
\vspace{-0.1cm}
\begin{equation}
\alpha_{i,j,t}^u=\frac{\textbf{\textit{exp}}(\tau\cdot W_Q^uh_i^t\cdot(W_K^uh_j^t)^\mathrm{T})}
{\sum_{k\in \mathcal{A}_{+i}^t}\textbf{\textit{exp}}(\tau\cdot W_Q^uh_i^t\cdot(W_K^uh_k^t)^\mathrm{T})},\label{eq9}
\end{equation}
where $\tau$ is a scaling factor. Next, the outputs of $U$ attention heads for node $i$ are concatenated and then fed into function $ \bm{\sigma}$ to produce the output of the convolutional layer, expressed as $h_i^{\prime t}= \bm{\sigma}(\textbf{\textit{con}}[\sum_{j\in \mathcal{A}_{+j}^t}a_{i,j,t}^uW_V^uh_j^t, \forall u])$, where \textbf{\textit{con}} denotes the concatenation of the outputs of $U$ attention heads. Note that, more graphical information can be extracted through increasing the number of 
convolutional layers.

\vspace{-0.1cm}
\subsection{Parameter Updating}
Similar to the traditional deep Q network (DQN) algorithm using Q values to estimate the long-term cumulative reward of taking a certain action at the current state, the MGARL algorithm utilizes both the training network and the target network, where the training network parameters are updated by optimizing the loss function of the target network. Through extracting graphical information to facilitate cooperative learning, the MGARL method can better capture the dynamics and irregularities than the traditional DRL methods. In order to utilize historical data for training and improve sample utilization, the tuple $ \left\langle o_i^t,a_i^t,r_i^t,o_i^{t+1}, \mathbf{C}_i^t, \mathbf{C}_i^{t+1}\right\rangle $ of vehicular agent $i$ is stored in the shared replay buffer in each TS. When training, $S$ sets of samples are randomly selected to update the parameters of the training network. Particularly, in order to achieve stable interactions between agents, MGARL optimizes the attention weight distribution of the last convolutional layer to minimize the Kullback-Leibler divergence of the attention weight distribution in the current TS over the weights in the next TS. Then, the loss function of the target network  is expressed as
\vspace{-0.1cm}
\begin{equation}
\begin{aligned}
 \mathcal L(\theta)=\frac1S\sum_S\frac1M
\sum_{i=1}^M(y_i^t-Q_i(\mathcal{O}_{i,\mathbf{C}_i^t},a_i^t;\theta))^2+\\
\vspace{-0.3cm}
\lambda\frac1U\sum_{u=1}^UD_{KL}(  \mathcal G_{u,t}^k(\mathcal{O}_{i,\mathbf{C}_i^t};\theta)\Vert(\mathcal G_{u,t+1}^k(\mathcal{O}_{i,\mathbf{C}_i^{t+1}};\theta)), \label{eq12}
\end{aligned}
\vspace{-0.15cm}
\end{equation}

\noindent where $y_i^t=r_i^t+\gamma {max}_{a^{\prime}}Q_i^{\prime}(\mathcal{O}_{i,\mathbf{C}_i^{t+1}},a_i^{\prime};\theta^{\prime}))^2$ is the Q value generated by the target network with parameter $\theta^{\prime}$. $\gamma$ is the discount factor. $\mathcal{O}_{i,\mathbf{C}_i^t}$ denotes the set of observations of nodes in agent $i$'s local region determined by adjacency matrix $\mathbf{C}_i^t$. Q function parameterized by $\theta$ takes $\mathcal{O}_{i, \mathbf{C}_i^t}$ as input and outputs $Q$ value for agent $i$. $\lambda$ is the coefficient for the loss and $\mathcal G_{u,t}^k(\mathcal{O}_{i,\mathbf{C}_i^t};\theta)$ denotes the attention weight distribution of relation representations of attention head $u$ in convolutional layer $k$ for agent~$i$. The details of the training process for the proposed MGARL-based algorithm are listed in Algorithm 1.

%\subsection{Algorithm Design}

%As shown in Algorithm 1, the details of the proposed MGARL-based content caching-assisted VEC scheme include the following steps:
%\begin{itemize}
  %\item [1)]  
  %At the beginning of each episode, model the vehicular network topology as a graph $\mathcal{G}=\{\mathcal{V},\mathcal{E}\}$. Construct the adjacency matrix $\mathbf{C}_i^t$ for all agents in each time slot and initialize the parameter $\theta_i$ of the Q network maintained by agent $i$;      
  %\item [2)]
  %Each VU completes one task in one time slot by fetching the result content from SNs, task offloading, or local computing; 
  %\item [3)]
  %Each VU gets its observation $o_i^t$ including its own location, its own caching state, the current remaining storage capacity of all SNs;
  %\item [4)]
  %Each VU makes its content caching decision and performs the action $a_i^t$, and then obtains the system reward $r_i^t$;
  %\item [5)]
  %Each VU stores the tuple $ \left\langle o_i^t,a_i^t,r_i^t,o_i^{t+1}, C_i^t, C_i^{t+1}\right\rangle $ in the shared replay buffer;
  %\item [6)]
  %A fully connected network encodes $o_i^t$ to generate the node feature $h_i^t$. Then the convolutional layer integrates the features of $\mathcal{A}_{+i}$ and generates the latent feature $h_i^{\prime t}$ by adopting multi-head  attention as the convolution kernel;
  %\item [7)]
  %The parameter updating module updates the parameter $\theta_i$ of Q network maintained by agent $i$ to minimize the loss function by Eq.\ref{eq12};
  %\item [8)]
  %Repeat the above steps until the algorithm converges.
%\end{itemize}

%\vspace{-0.2cm}
\subsection{Complexity Analysis}
Let us now discuss the complexity of the graph attention convolutional layer in TS $t$ including the feature mapping of nodes and the calculation of attention weights. We assume that the constructed graph in TS $t$ consists of $ \lvert \mathcal{E}_t \rvert $ edges and each node feature is mapped from dimension $F$ to space in dimension $ F'$. Then, the computational complexity for mapping node features is represented as $\mathcal O\left(MFF'\right)$ while the computational complexity in calculating attention weights is related to the number of edges, which is expressed as $\mathcal O\left(\lvert \mathcal{E}_t \rvert F' \right)$. Therefore, the total computational complexity with $U$ attention heads is determined by the number of VUs, the number of edges, and the node feature dimension, given by $ \mathcal O\left(U(MFF'+ \lvert \mathcal{E}_t \rvert F' )\right)$.

\section{Simulation Results}
\subsection{Simulation Settings}
We consider a Manhattan vehicular network model, where the length and width of the road are both $1$~km. By default, we set $M=20$, $L=16$, $d_k\in \left [ 50,80 \right ]$~MB, $b_k\in \left [ 6,16 \right ]$~MB, $s_k=50$ Megacycles $(\forall k)$, $B=10$~MHz, 
%$P_m=23$~dBm $(\forall m)$ and the transmission power of RSUs is set to $40$~dBm, 
$\sigma^2=-174$~dBm/Hz, $\delta=0.5$, $re_m=2\ (\forall m)$, and $pl \in \{0, -0.5\}$. The transmit power of RSUs and VUs are set to $40$~dBm and $23$~dBm, respectively. The caching capacity of RSUs and VUs are $100$~MB and $50$~MB, respectively. The computation capability of RSUs and VUs are $1$~GHz and $0.8$~GHz, respectively. The communication range of RSUs and V2V communication are set as $400$~m and $300$~m, respectively. We adopt the channel model according to~\cite{forexecution2}. In terms of the VU mobility model, we set $\eta=0.4$ and the mean range of VUs' velocity as $[36, 54]$~km/h, and other parameters setting please refer to~\cite{mobility}. With regard to the learning configurations, $\gamma=0.9$, $S=128$, the learning rate is $0.001$ and the buffer capacity is $2000$. For fair comparison, we choose the proposed w/o attention based scheme, multi-agent Independent Double
Deep Q Network (IDDQN) based scheme and the random content caching scheme.

%\begin{table}
%\vspace{-0.1cm}
  %\begin{center}
%\renewcommand{\arraystretch}{1.3}
    %\caption{\textsc{SIMULATION PARAMETERS}}
    %\vspace{-0.2cm}
    %\begin{tabular}{p{5cm}|p{3cm}} 
       
    %\hline \hline
      %Description &  Value  \tabularnewline
      %\hline
     %\rule{-2.5pt}{1pt} Path loss &\cite{b18}\tabularnewline
     %\rule{-2.5pt}{1pt} Transmission power of vehicle users & 30 dBm \tabularnewline
     %\rule{-2.5pt}{1pt} Transmission bandwidth  & 10 MHz  \tabularnewline
     %\rule{-2.5pt}{1pt} Additive White Gaussian noise power &-174 dBm/Hz \tabularnewline
     %\rule{0pt}{1pt}Number of CPU cycles required to process each  task   & 50 Megacycles \tabularnewline
     %\rule{0pt}{1pt}Total content types    &  10 \tabularnewline
     %\rule{-2.5pt}{1pt} The parameter of Zipf Distribution $\delta$  & 0.5 \tabularnewline
     %\rule{0pt}{1pt}Computation capability of RSUs &$1000$~Hz \tabularnewline
     %\rule{-2.5pt}{1pt} Computation capability of VUs  & $800$~Hz\tabularnewline
     %\rule{-2.5pt}{1pt} Learning rate & 0.05   \tabularnewline
     %\rule{-2.5pt}{1pt} Discount rate  & 0.9   \tabularnewline
     %\rule{-2.5pt}{1pt} Buffer capacity & 2000 \tabularnewline
     %\rule{-2.5pt}{1pt} Size of mini-batch S  & 128   \tabularnewline
       % \hline \hline
    %\end{tabular}
  %\end{center}
  %\vspace{-0.3cm}
%\end{table}

\subsection{Performance Evaluation}

\begin{figure}
\vspace{-0.3cm}
\centering 
\includegraphics[width=0.9\linewidth]{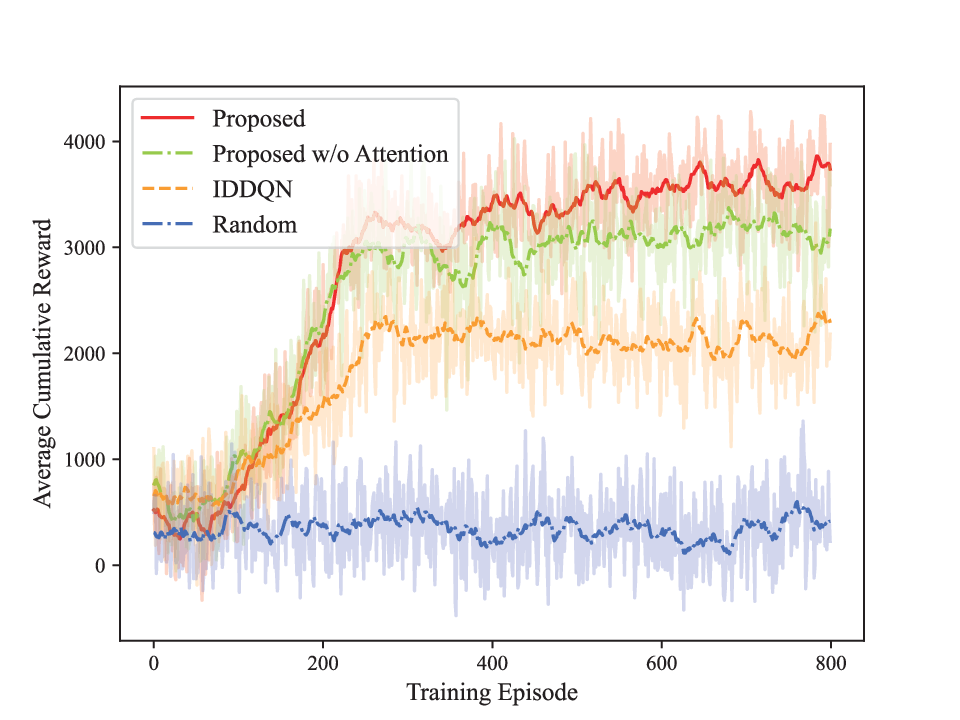}
\vspace{-0.3cm}
    \caption{Comparison of the convergence.}
    \label{converge}
\vspace{-0.4cm}
\end{figure}
Fig.~\ref{converge} presents the convergence performance of all schemes. It can be observed that the cumulative reward of the proposed scheme is higher than the baselines. This is attributed to the fact that the proposed scheme, in conjunction with the proposed w/o attention and IDDQN-based scheme, can learn policy by continuously interacting with the environment. Moreover, the proposed scheme utilizes graph attention convolution kernels to capture the relationships among agents to further enhance cooperation, which is more adaptable to irregular network topologies and consequently has the highest cumulative reward.

\begin{figure}
\vspace{-0.1cm}
    \centering
\includegraphics[width=0.9\linewidth]{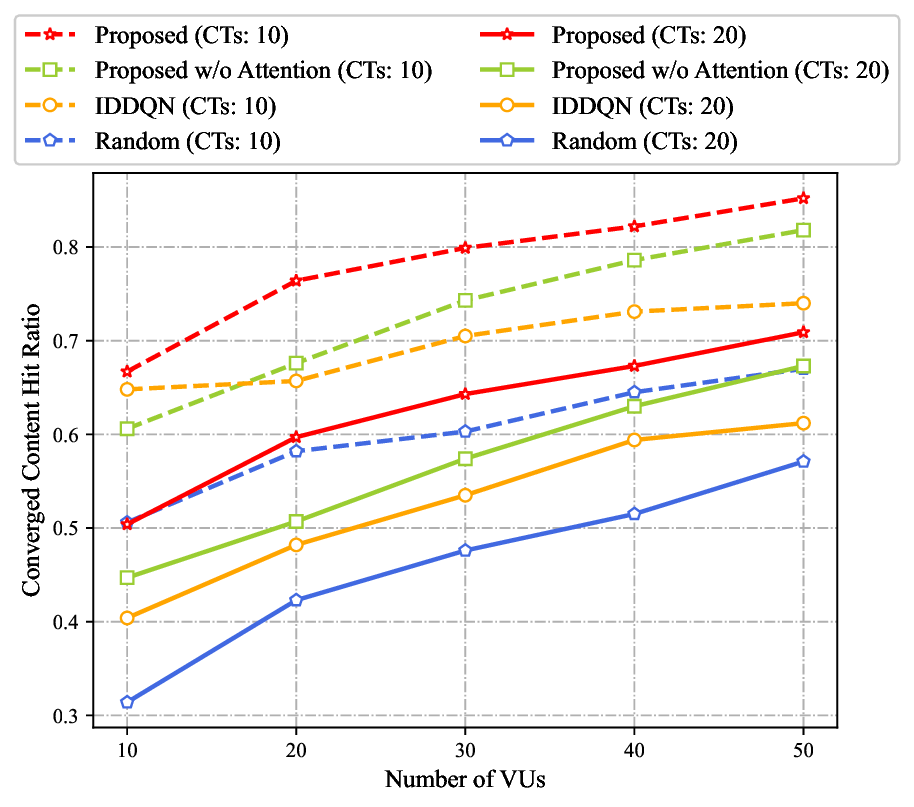}
    \caption{The converged content hit ratio versus the number of VUs.}
    \label{hit}
\vspace{-0.2cm}
\end{figure}

\begin{figure}
    \centering
    \vspace{-0.2cm}
\includegraphics[width=0.9\linewidth]{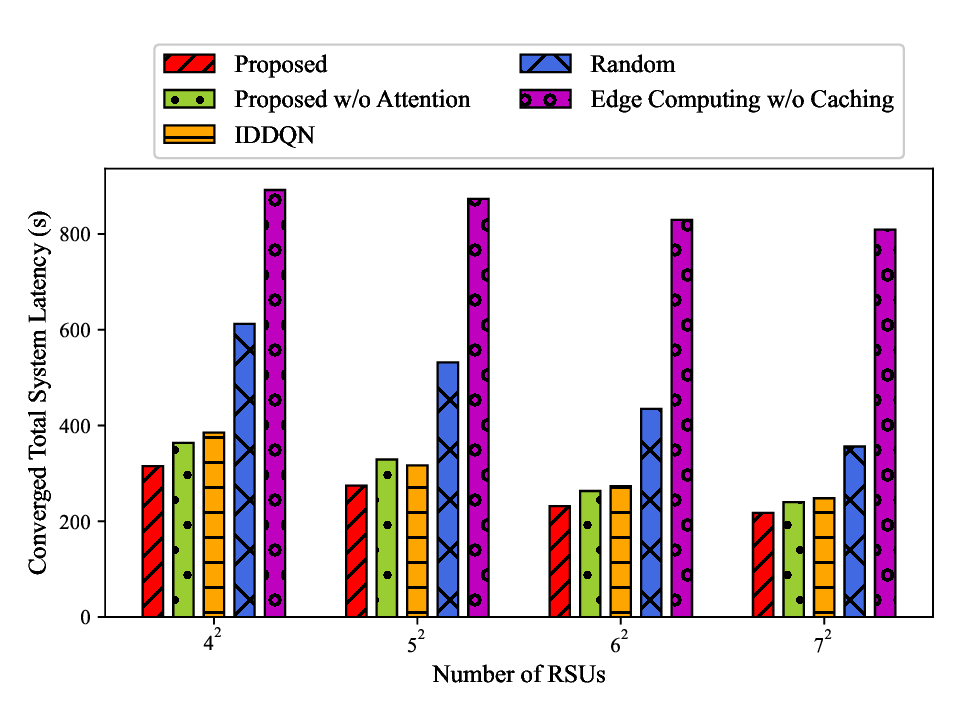}
    \caption{The converged total system latency versus the number of RSUs.}
    \label{latency}
\vspace{-0.5cm}
\end{figure}

The converged content hit ratio versus the number of VUs is investigated in Fig.~\ref{hit} when the number of content types (CTs) is 10 and 20. The first point to observe is that the converged content hit ratio increases as the number of VUs increases. This is because more users lead to the increasing number of content requests, thus increasing the probability of reusing the same cached content. Secondly, it is clear that the content hit ratio is reduced when the number of CTs is increased from $10$ to $20$. The underlying reason is that due to the limited caching capacity of SNs, they may not cache all the contents, which decreases the hit ratio of each content. Moreover, we can see that our proposed scheme outperforms other schemes in improving the content hit ratio regardless of both the number of CTs and the number of VUs, which verifies the effectiveness of the proposed scheme in improving caching resource utilization.

Fig.~\ref{latency} compares the converged total system latency versus the number of RSUs. Firstly, as the number of RSUs increases, the converged total system latency exhibits a downward trend. The underlying reason for this phenomenon is that with more RSUs having computation and caching capability, VUs are more likely to fetch the cached result content from the surrounding RSUs, thus avoiding the repeated uploading and computing process. Secondly, it can be seen that the proposed scheme outperforms other schemes under different numbers of RSUs, since multiple graph attention convolutional layers can extract more hidden structural information to facilitate cooperative learning. This phenomenon implies that the proposed scheme can make better use of densely deployed cache resources to further reduce the total system latency.
%We observe that, the proposed scheme the increase in caching resources for SNs improves the cumulative reward. This is because more caching resources facilitate more task contents to be cached, which ultimately reduces long-term task computing latency and results in a higher reward. 
%\textcolor{blue}{It is worth mentioning that the proposed scheme has the capacity to acquire optimal performance even with limited caching resources due to the efficient use of which to reduce long-term task computing latency. }

\section{Conclusion}

In this paper, we proposed a content caching-assisted VEC framework by taking into account the reuse of popular task computing results. To adapt to the irregular network topology and the environmental uncertainty, we developed an MGARL-based edge caching scheme for VEC networks by utilizing the cooperation among agents with the integration information of neighboring nodes in decision-making. Compared to the baselines, the proposed scheme can better learn the irregular topology dynamics, thus significantly reducing the task computing latency and improving the utilization of densely deployed caching resources.

\bibliographystyle{ieeetr}
\bibliography{reference}

%\bibitem{b12} L. T. Tan and R. Q. Hu, ``Mobility-Aware Edge Caching and Computing in Vehicle Networks: A Deep Reinforcement Learning," \textit{IEEE Trans. Veh. Technol.}, vol. 67, no. 11, pp. 10190-10203, Nov. 2018.

\end{document}